# NORAD Tracking of the February 2022 Starlink Satellites (and the Possible Immediate Loss of 32 Satellites)

Fernando Guarnieri[1], Bruce T. Tsurutani[2], Rajkumar Hajra[3], Ezequiel Echer[4] and Gurbax S. Lakhina[5]

[1]Rua Serra das Vertentes, 56, São José dos Campos, SP, Brazil

[2]Retired, Pasadena, California USA

[3] CAS Key Laboratory of Geospace Environment, School of Earth and Space Sciences, University of Science and Technology of China, Hefei, China

[4]Brazilian National Institute for Space Research – INPE, Brazil

[5]Vashi, Navi Mumbai 400703, India

**Abstract**

The North American Aerospace Defense Command (NORAD) tracking of the SpaceX Starlink satellite launch on February 03, 2022 is reviewed. Of the 49 Starlink satellites released into orbit, 38 were eventually lost. Thirty-two of the satellites were never tracked by NORAD. There have been three articles written proposing physical mechanisms to explain the satellite losses. It is argued that none of the proposed mechanisms can explain the immediate loss of 32 of the 49 satellites. The non-availability of telemetry data from the lost satellites has hindered the search for a physical mechanism to explain the density increase observed in a very short time interval.

1. Introduction

Geomagnetic storms are caused by magnetic reconnection between southward interplanetary magnetic field (IMF) and the Earth's dayside magnetic fields (Dungey, 1961; Gonzalez et al., 1994). The reconnected magnetic fields and solar wind plasma are convected to the midnight sector of the Earth's magnetosphere where the magnetic fields are reconnected again. The reconnected fields and plasma are jetted from the magnetotail towards the inner magnetosphere, causing auroras in the midnight sector at geomagnetic latitudes of 65° to 70° and slightly lower (the auroras occur both in the northern and southern polar regions). The auroras also spread to all longitudes covering the Earth's magnetosphere at the above latitudes if the storm is intense and long lasting.

The auroras are caused by the influx of energetic ~1 to 30 keV electrons into the auroral zone atmosphere (Anderson, 1958; Carlson et al., 1998; Hosokawa et al., 2020) impacting atmospheric atoms and molecules at a height of ~110 to 90 km. The excited atoms and molecules decay emitting violet, green and red light. The influx of the energetic electrons are believed to also causes the upwelling of oxygen ions to heights where they will impact the orbiting satellites, causing enhanced drag on the satellites and eventual lowering of their orbits. This is the standard picture of low altitude satellite drag during magnetic storms.

2. Starlink Launch

On February 03, 2022, at 18:13 UT, SpaceX launched the rocket Falcon 9 Block 5 with the objective of deploying the satellites for the Starlink Group 4-7, the sixth launch to the Starlink Shell 4. This launch received the international COSPAR identification ID 2022-010. A video by

Manley (2021) illustrates how two stacks of Starlink satellites could be put into orbit from a single launch vehicle. In the example each stack of ~30 satellites can be released in different directions. When the satellites separate from this stack, they start individual movements, sometimes colliding gently with others before enter into their individual flight orbits. For the February 3 launch, there were 20 satellites in each stack. After the launch, the satellites may be put into edge-on directions with the solar panels parallel to the satellite bodies to attempt to reduce drag. However, a telecommand is necessary to make them keep the safehold strategy, demanding some time and requiring some minimum antenna pointing.

The SpaceX mission under analysis was composed of 49 Starlink satellites that were initially planned to orbit the Earth at ~540 km circular low-Earth orbit (LEO). The initial planned elliptical orbit was 338 km x 210 km, at an inclination of 53.22°. Once the initial elliptical orbits were obtained, SpaceX would use onboard propulsion to raise the orbits.

The February 3, 2022 deployment of the satellites occurred 15 minutes and 31 seconds after the liftoff, at a release time of 18:28 UT. SpaceX considered the launch successful, since the releasing of the satellites occurred in the expected orbits, the rocket stage was recovered as planned, and all the satellites were able to switch to autonomous flight.

3. Space Weather for the Period

From the time of launch until a day after it, the near-Earth space weather conditions were disturbed with the occurrences of two geomagnetic storms. Figure 1 shows the interplanetary conditions for the period as well as the geomagnetic SYM-H index.

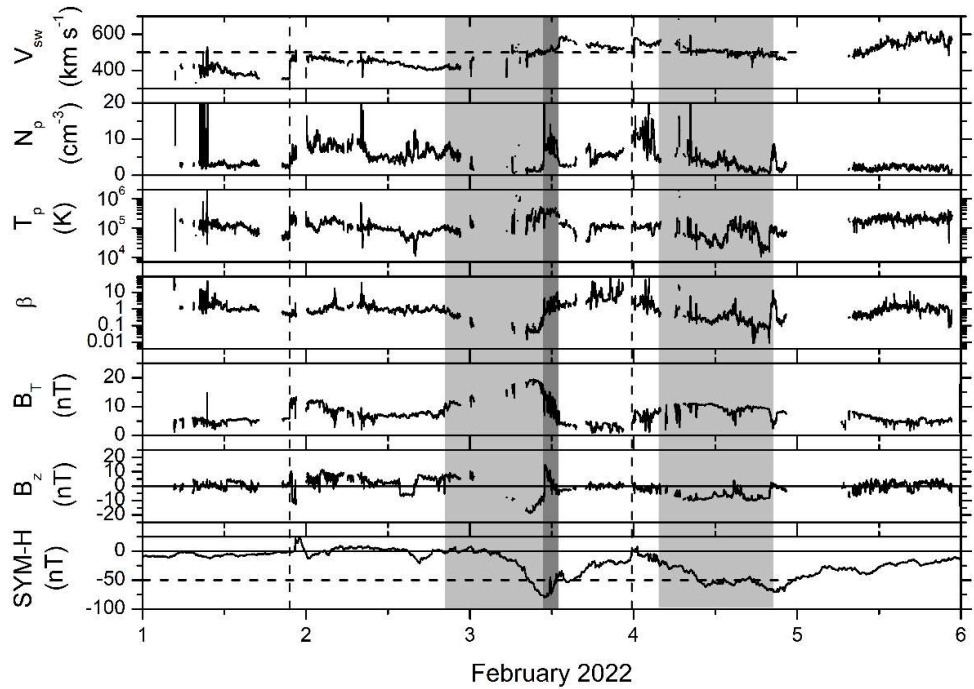

Figure 1 - The interplanetary and geomagnetic conditions during 1–5 February 2022. The interplanetary data at 1 AU are time shifted from the spacecraft location at the L1 libration point ~0.01 AU upstream of the Earth to the nose of the Earth's bow shock. From top to bottom are: the solar wind speed ($V_{sw}$), the plasma density ($N_p$), temperature ($T_p$), plasma-$\beta$, the interplanetary magnetic field (IMF magnitude ($B_T$), and the $B_z$ component. The bottom panel is the SYM-H index. The IMF component is given in the geocentric solar magnetospheric (GSM) coordinate system. The interplanetary data were obtained from the Goddard Space Data System (https://omniweb.gsfc.nasa.gov/) and the storm-time SYM-H index from the Kyoto University data system (https://wdc.kugi.kyoto-u.ac.jp/). The vertical dashed lines indicate interplanetary shocks. The light gray shadings indicate magnetic clouds (MCs), and the dark gray shading indicates a solar filament propagated to 1 AU. Taken from Tsurutani et al. (2022)

A few days prior to the Starlink satellite launch, on January 29, at ~23:00 UT, a M1.0 solar flare erupted from the active region 12936. A coronal mass ejection (CME) was released from this same active region at 23:36 UT, and its upstream shock reached the Earth on February 02, 2022, at 21:34 UT. The geomagnetic effect of an interplanetary CME (ICME) impacting the Earth's magnetosphere was the occurrence of a moderate storm (Gonzalez et al., 1994; Echer et al., 2008) with peak intensity of SYM-H= -80 nT at 10:56 UT on 3 February. A second CME impacted the Earth causing a second (moderate) magnetic storm of SYM-H = -71 nT at 20:59 UT on February 04.

In Figure 1 the vertical dashed line at ~21:34 UT on 1 February indicates a fast forward shock arrival. The shock causes a sudden impulse (SI$^+$) of 22 nT noted in the SYM-H index. The high-density sheath is present from the shock to a magnetic cloud (MC) portion of the ICME. The sheath did not contain major southward IMFs, so was generally not geoeffective. The MC portion of the ICME is identified by high IMF magnitude ($B_T$) and low plasma-β (the ratio between the plasma thermal pressure and the magnetic pressure; Burlaga et al., 1981; Tsurutani et al., 1988) and is shown by light gray shading. The MC extends from ~21:00 UT on 2 February to ~13:11 UT on 3 February. The IMF Bz component of the MC has the characteristic "fluxrope" configuration with a southward component followed by a northward component. When IMF is southward, the symmetric ring current index SYM-H decreases to a peak value of -80 nT at ~10:56 UT on 3 February. Thus, the magnetic storm is caused by the magnetic reconnection process as postulated by Dungey (1961). The dark gray shaded region is the high-density solar filament portion of the ICME (Illing & Hundhausen, 1986; Burlaga et al., 1998). The impact of the filament causes a compression of the magnetosphere and a sudden increase in the SYM-H index to -39 nT.

The speed of the CME at 1 AU was approximately 500 km s$^{-1}$. This is classified as a moderately fast CME (faster than the slow solar wind speed of ~350 to 400 km s$^{-1}$), thus causing formation of the upstream shock and sheath. From this figure it is clear that SpaceX launched their Starlink satellites into a moderate intensity magnetic storm.

There is a second fast CME which occurred about one day later, although the exact solar origin location and release time was not analyzed. The shock is identified by a vertical dashed line at ~00:04 UT on 4 February. The shock caused a SI$^+$ of intensity ~17 nT. The sheath of the second event did not contain major IMF southward component, so again it was not geoeffective. The MC portion of the second ICME is indicated by a light gray shading from ~02:25 UT to ~20:12 UT on 4 February. The MC had a peak IMF $B_T$ of 11 nT at ~10:59 UT. The MC $B_z$ component profile is different from the previous MC. $B_z$ is negative or zero throughout the MC. The negative $B_z$ causes the second magnetic storm of peak intensity -71 nT at ~20:59 UT on 4 February. There was no solar filament during this second ICME event.

The effects of these storms on the atmospheric mass density was analyzed using data from the Swarm B satellite. Swarm B is in a circular orbit at ~500km, with an inclination of ~88.0° and orbital period of ~90 minutes, so there are about 15 orbits per day. The orbits have been numbered for each day. In Figure 2 at 00 UT 2 February, the satellite was at ~ -10° latitude at ~09 local time (LT) on the dayside, and was moving towards the south pole. The mass density is ~3.5×10$^{-13}$ kg m$^{-3}$ (a light blue color). Continuing in time as the orbit crosses over the south pole and enters the nightside ionosphere at ~20 LT, it is noticed that between -53° and +53° the density reduces to ~1.5×10$^{-13}$ kg m$^{-3}$ (a dark blue color). The other orbits throughout 2 February show a similar pattern between the nightside and dayside passes.

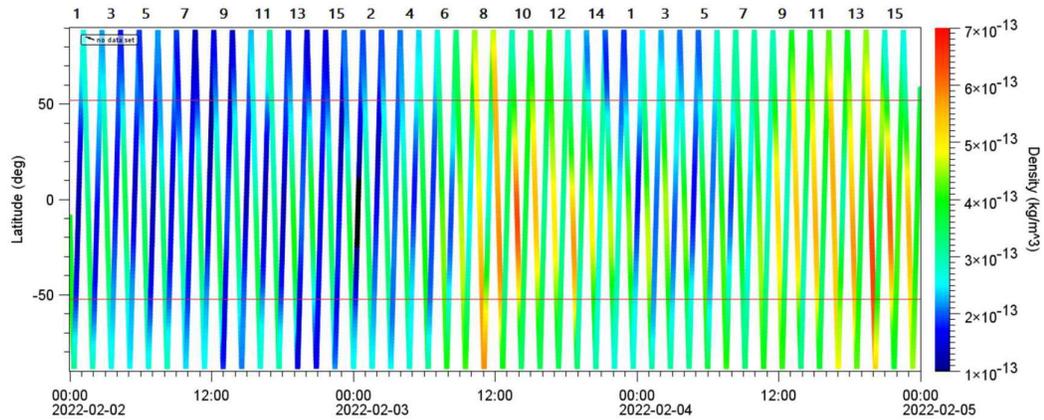

Figure 2. The Swarm B mass impact data for 2-4 February 2022. The mass density is shown as a function of UT (x-axis) and geographic latitude (y-axis). It can be noted that the observations cover both day (north-to-south hemispheric passes) and night (south-to-north hemispheric passes) sides of the globe. 2 February was a quiet day before the two magnetic storms and is shown as a "quiet-day reference". The mass density values are given in linear color scale on the right. Two red horizontal lines at +53° and -53° indicate the upper limits of the intended Starlink satellite orbits. Swarm B orbits on each day from the north pole to the south pole and back are marked by numbers from 1 to 15. Partial orbit 1 for 2 February is shown at the beginning of the figure.

On orbit 8 of 3 February, there is the first sign of a change (increase) in the mass impact at middle and low latitudes (~$5.0\times10^{-13}$ kg m$^{-3}$, an orange color). This occurs at the south pole crossing at ~10 UT, just before the peak of the first magnetic storm. There is a density enhancement (orange coloration) throughout this downward dayside pass, across the magnetic equator and to the south pole. There is a local maximum of density at ~14 UT and ~09 LT at 10° latitude. On orbits 9–13 of 3 February, the predominant density enhancements are on the dayside passes in the equatorial and midlatitude ranges. The enhancements are larger than those at higher latitudes. The maximum density of ~$5.5\times10^{-13}$ kg m$^{-3}$ occurred at ~19 UT and ~09 LT. This represents a density peak increase of ~50% relative to the quiet daytime density (2 February).

On orbits 9–13 of 3 February, the nightside equatorial and midlatitude densities are ~$3.5\times10^{-13}$ kg m$^{-3}$ (light green color). This is higher than the 2 February (quiet time) nightside densities of ~$1.5\times10^{-13}$ kg m$^{-3}$. Thus, during the magnetic storm, the nightside peak densities increased by ~100%. It is noted that the nighttime peak densities are less than the daytime peak densities. This latter feature will be explained in the Discussion section.

The high impact mass (orange color) fades out by the end of 3 February and does not start again until orbit 8 of 4 February. A density peak of ~$5.3\times10^{-13}$ kg m$^{-3}$ at the equatorial region on orbit 8 occurred at ~1200 UT. This is approximately 10 hours after the slowly developing second magnetic storm started at ~ 0015 UT on 4 February. From orbit 8 to 11 the predominant density enhancement occurs at the equatorial to middle latitudes with little or no enhanced impact in the auroral/polar regions. The maximum density of ~$6.3\times10^{-13}$ kg m$^{-3}$ occurred at 20:00 UT on dayside pass 14, and extended from ~-15° to -60° latitudes. The peak time is coincident with the peak in the second magnetic storm. On passes 15 and 16, the density decreases, and the enhanced density occurs mainly at the equator and middle latitudes. The maximum density during this second storm event was ~80% higher than the dayside density values detected on 2 February.

The nightside pass density on orbit 14 on 4 February was ~$4.3\times10^{-13}$ kg m$^{-3}$. This is ~190% higher than the quiet time value on 2 February. The nighttime peak densities are lower than the daytime peak densities, similar to the first storm features. The data for 5 and 6 February look similar to the quiet day interval of 2 February, so are not shown to conserve space.

## 4. Magnetic Storm Effects on Starlink Satellite Survivability

Among the 49 released satellites, only 17 could be tracked by the North American Defense Command (NORAD) some days later. Thirty-two satellites were never listed by NORAD, thus we assume that they were immediately lost after launch. This may have happened due to problems in tracking them (due to extremely fast orbital decays in the first hours after the release or due to substantially different satellite positions than expected for the launch).

This Starlink loss is a significant reduction in the Starlink launch efficiency, which had been 97% successful for the last 34 launches. For this specific launch, only 22.45% of the load survived and became operational. Table 1 shows the statistics of the launches for the Starlink Satellites version 1.5.

| Mission Number | Mission Name | Deployed | Working | Efficiency |
|---|---|---|---|---|
| 30 | Group 2-1 | 51 | 50 | 98.04% |
| 31 | Group 4-1 | 53 | 52 | 98.11% |
| 32 | Group 4-3 | 48 | 48 | 100.00% |
| 33 | Group 4-4 | 52 | 51 | 98.08% |
| 34 | Group 4-5 | 49 | 49 | 100.00% |
| 35 | Group 4-6 | 49 | 49 | 100.00% |
| 36 | Group 4-7 | 49 | 11 | 22.45% |
| 37 | Group 4-8 | 46 | 46 | 100.00% |
| 38 | Group 4-11 | 50 | 50 | 100.00% |
| 39 | Group 4-9 | 47 | 47 | 100.00% |
| 40 | Group 4-10 | 48 | 47 | 97.92% |
| 41 | Group 4-12 | 53 | 47 | 88.68% |
| 42 | Group 4-14 | 53 | 53 | 100.00% |
| 43 | Group 4-16 | 53 | 53 | 100.00% |
| 44 | Group 4-17 | 53 | 53 | 100.00% |
| 45 | Group 4-13 | 53 | 53 | 100.00% |
| 46 | Group 4-15 | 53 | 53 | 100.00% |
| 47 | Group 4-18 | 53 | 53 | 100,00% |
| 48 | Group 4-19 | 53 | 53 | 100.00% |
| 49 | Group 4-21 | 53 | 53 | 100.00% |
| 50 | Group 3-1 | 46 | 46 | 100.00% |
| 51 | Group 4-22 | 53 | 53 | 100.00% |

| 52 | Group 3-2 | 46 | 46 | 100.00% |
|---|---|---|---|---|
| 53 | Group 4-25 | 53 | 51 | 96.23% |
| 54 | Group 4-26 | 52 | 51 | 98.08% |
| 55 | Group 3-3 | 46 | 46 | 100.00% |
| 56 | Group 4-27 | 53 | 53 | 100.00% |
| 57 | Group 4-23 | 54 | 53 | 98.15% |
| 58 | Group 3-4 | 46 | 46 | 100.00% |
| 59 | Group 4-20 | 51 | 51 | 100.00% |
| 60 | Group 4-2 | 34 | 34 | 100.00% |
| 61 | Group 4-34 | 54 | 54 | 100.00% |
| 62 | Group 4-35 | 52 | 52 | 100.00% |
| 63 | Group 4-29 | 52 | 52 | 100.00% |
| 64 | Group 4-36 | 54 | 54 | 100.00% |
|  | TOTAL | 1765 | 1713 | 97.05% |

Table 1 – Statistics on Startlink satellite version 1.5 launches until October, 2022. Table compiled with data collected from Wikipedia Contributors, 2022.

On February 5, a first group of 6 Starlink satellite tracking was made available by NORAD. All these satellites had very low perigees, ~200 km altitude. The apogees were also very low, always below 350 km, and in some cases as low as 250 km. For the latter satellites, the apogees were not far from the perigee altitudes. Since the orbit injection velocities were too low for such unexpected low orbits, these satellites did not survive long. Some of these remaining 6 were also lost after a few tracking.

A second group of satellites, formed by 11 satellites, was tracked some days later, on February 8. These satellites were able to perform their ascending movements, changing from elliptical to circular orbits, and rising to higher and more stable intermediate orbits at ~350 km. The satellites were kept in this position for a few days. Afterwards their orbits were boosted to their final altitudes of ~540 km. These satellites were successfully saved and are now operational.

5. Satellite Tracking Timeline

In order to make it easier to understand all the sequence of events, a timeline was created with the space weather events, individual satellite tracking and other available information. Figure 3 shows this timeline. The satellites are identified by their NORAD numbers. Only those tracked after February 8 were linked to their Starlink numbers.

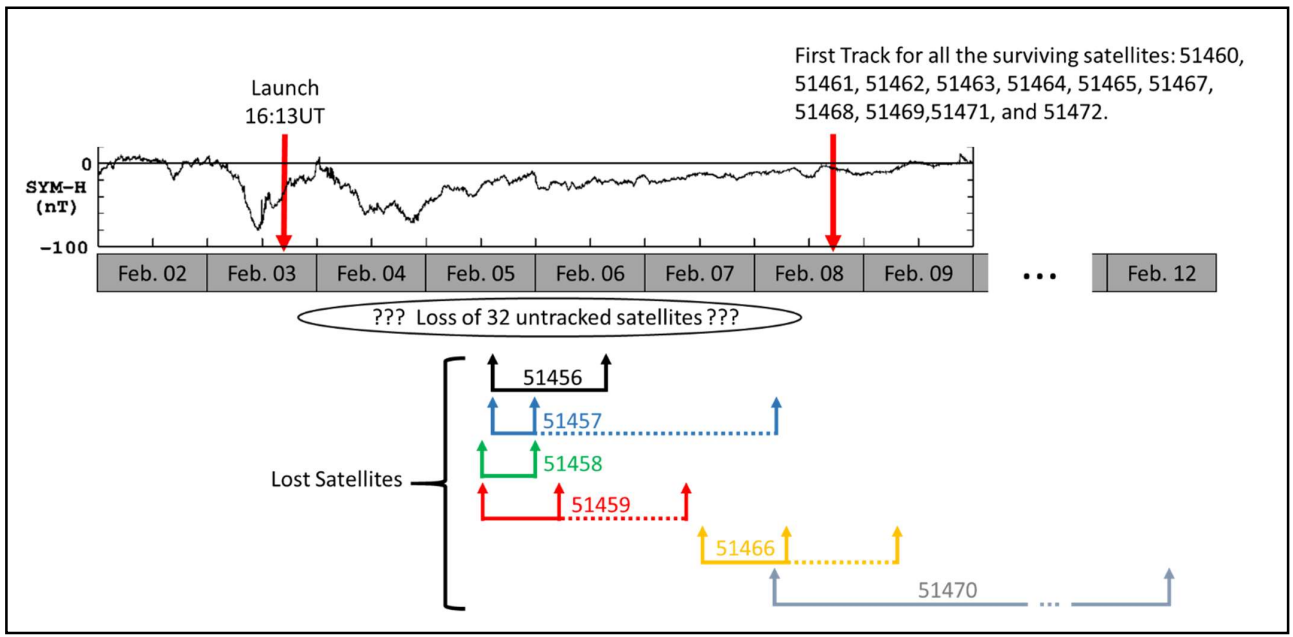

Figure 3 – Timeline for the satellite tracking occurring between February 2 and 12, 2022.

Figure 3 shows a plot of the SYM-H index which indicates the geomagnetic disturbances and the occurrence of geomagnetic storms. The two storm peaks are: SYM-H = -80 nT on Feb 3 and SYM-H = -71 nT on Feb 4. The red downward pointing arrows indicate the launch times, and the beginning of the tracking of the 11 surviving satellites, respectively. It should be noted that the Starlink satellites were launched in the recovery phase of the first storm (SYM-H increasing from its minimum value). Thus, the satellites are expected to have experienced effects from the first magnetic storm. It is also noticed that the second storm main phase started at the beginning of February 4 and continued for almost the entire day. Any Starlink satellites surviving the first storm would experience the effects of the second storm as well.

At the bottom of Figure 3, the upward arrows indicate the beginning and end times of the tracking for all other lost satellites (besides the original 32 satellites never tracked). For satellites 51457, 51459, and 51466, the extended dashed line and another arrow indicate the "official" decay times. An oval mark indicates the time interval when the 32 satellites were expected to be tracked, but were already lost.

6. Surviving Satellite Orbits Information

Figures 4 and 5 show the orbit information for the decayed satellites and for the operational satellites, respectively.

In Figure 4 the vertical axes give the satellite altitudes and the horizontal axes give the tracking sequences. The two dashed black lines indicates the perigees and the apogees expected for the satellite launches, 210 km and 338 km, respectively. The red and blue lines indicate the apogee and perigee at each tracking point. The date and time of the first and last tracking is indicated under the horizontal axis.

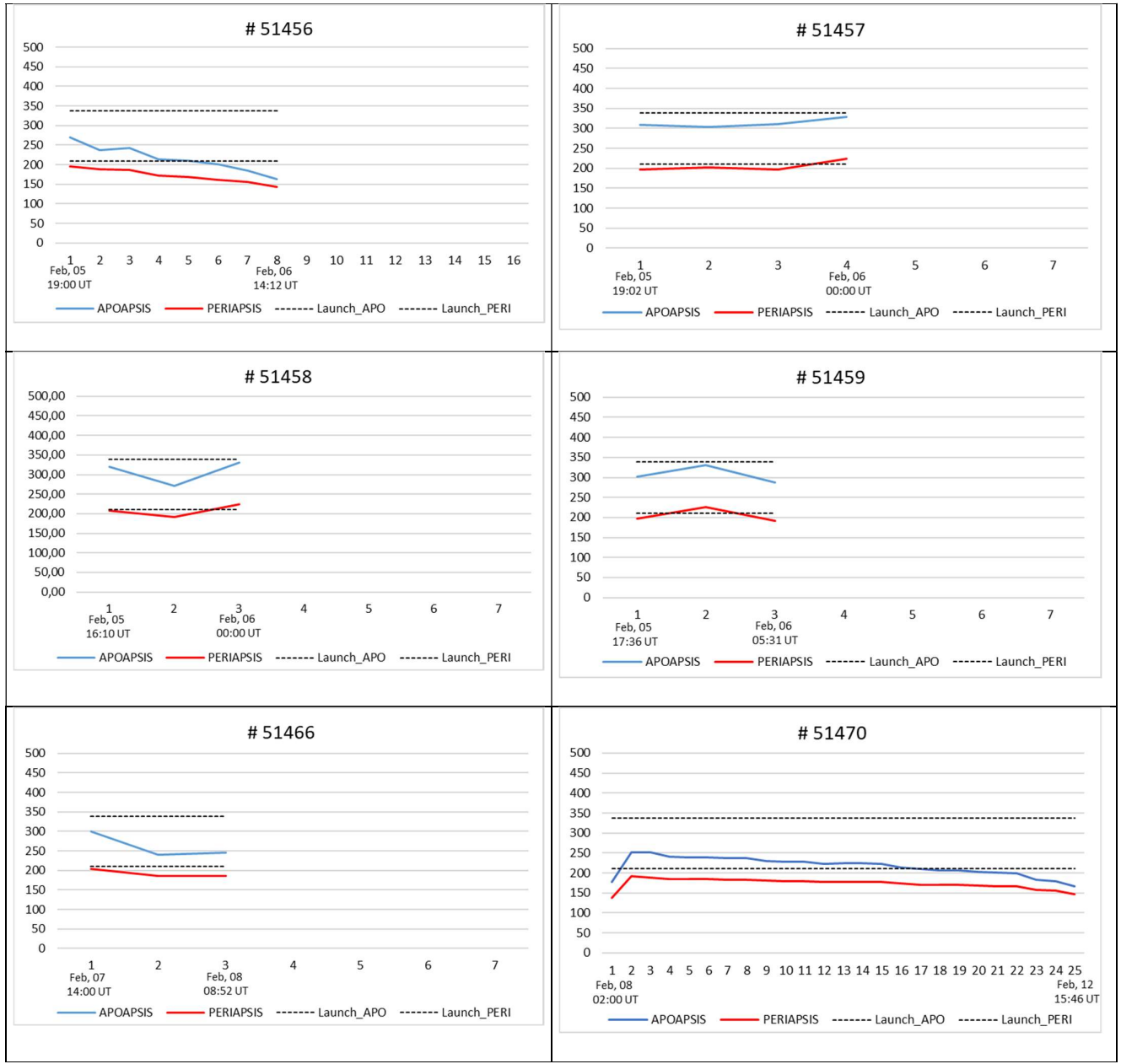

Figure 4 – Panels showing the orbit perigees (red lines) and apogees (blue lines) for each of the decayed satellites. The vertical axis in each panel gives the satellite height, and the horizontal axis indicates the tracking sequence. The dates and times under the horizontal axis indicates the time of the first and the last tracking. The two dashed black lines indicate the perigees and the apogees for the launch.

For all the above cases, the satellites were in very low orbits in the first track, close to the lowest orbits expected for the lowest perigees. The apogees were always very far (lower) from the expected values for the launch, and sometimes even closer to the values expected for perigees.

It can be noted that some satellites started to rise in altitude, but most likely were lost due to insufficient thrust in such low orbits with increased atmospheric drag.

A contrasting scenario is shown in Figure 5. All of these satellites survived the launching episode. After initial tracking by NORAD (all of them starting on February 8, 2022) they were boosted by onboard propulsion to safer (higher) altitudes.

The plots are in the same format as in Figure 4, but the horizontal axes now indicate the initial tracking (on February 8) until March 30.

It is interesting to note from Figure 5 that all of the satellites started their orbits in elliptical configurations, with apogee and perigee values much higher than the (decayed) satellites shown in Figure 4. The Figure 5 satellite orbits were very close to the specified values for the launch.

The orbit shapes changed to circular configurations (indicated by the merging of the red and blue lines) with subsequent altitude increases to intermediate orbital configurations. The rising to the final orbits were done very slowly, and none of the satellites had reached the final ~540 km altitude originally envisioned by March 30, almost two months after the launch.

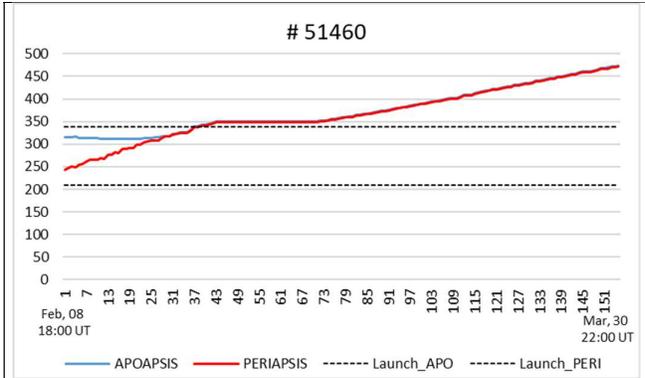
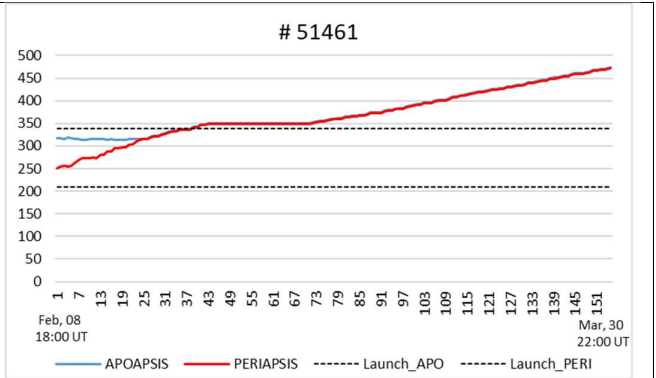
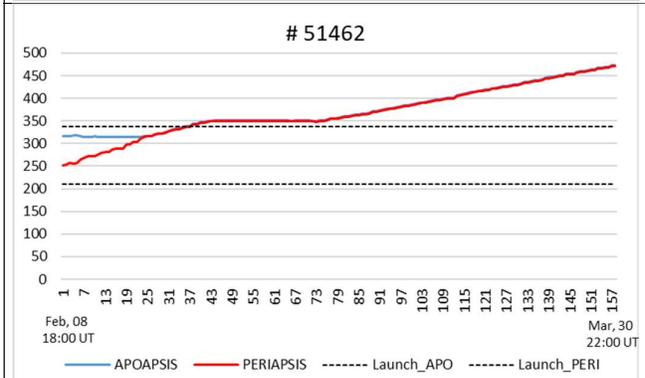
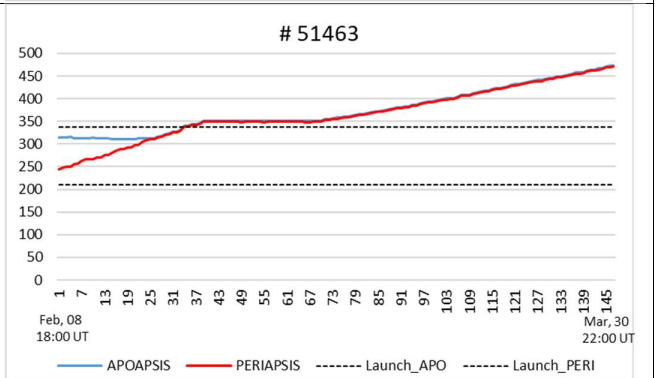
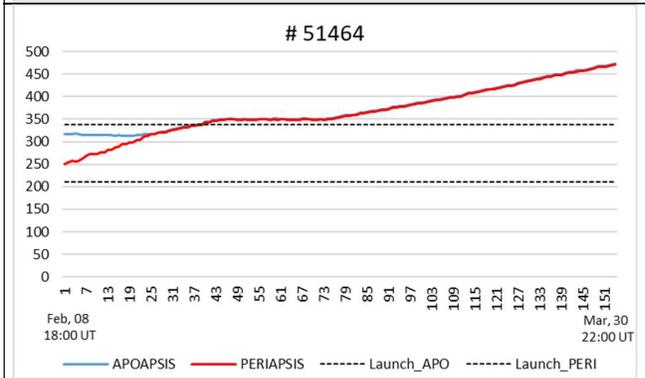
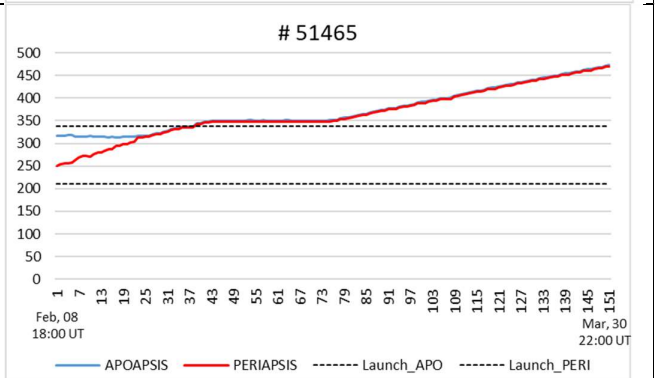
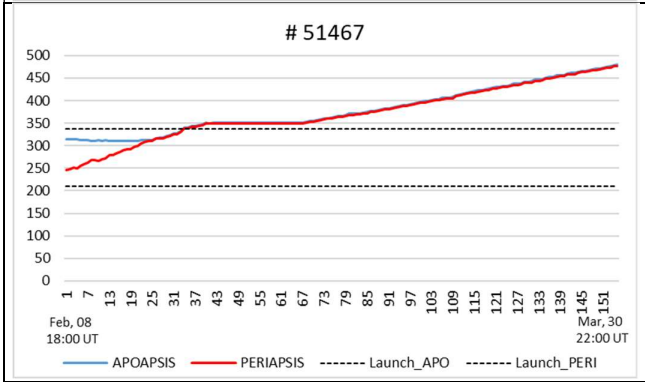
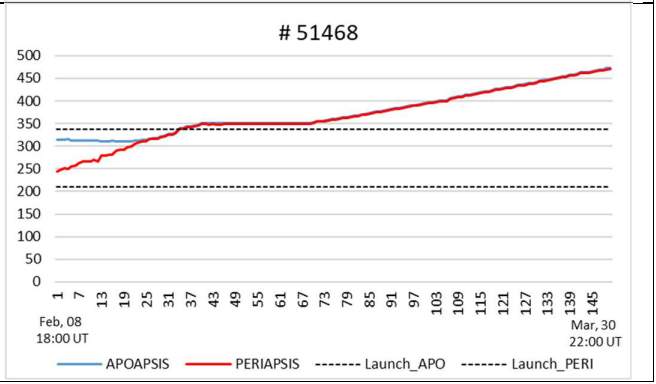

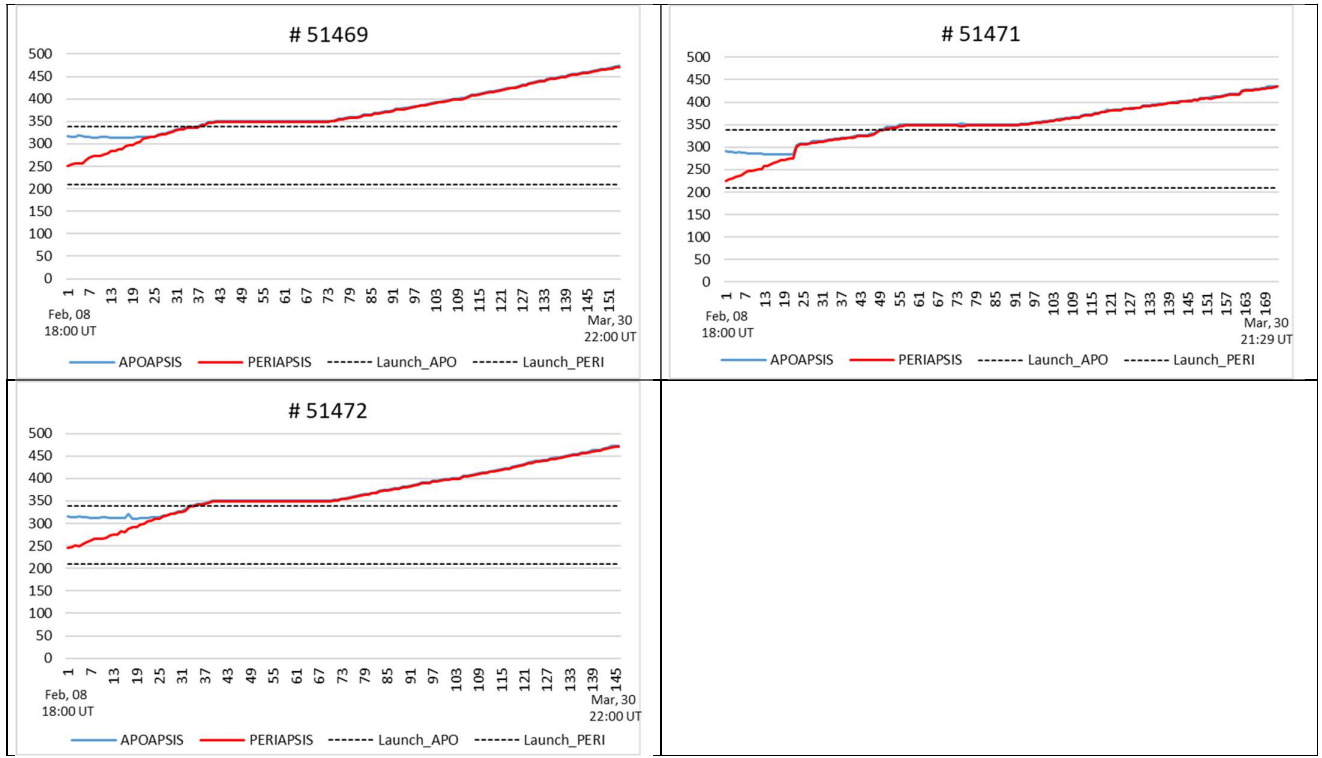

Figure 5 – Panels showing the orbit perigees (red lines) and apogees (blue lines) for the surviving satellites. The vertical axes give the satellite altitudes, and the horizontal axes indicate the tracking sequences from February 8 to March 30, 2022. The two dashed black lines indicate the perigees and apogees expected at the time of the launch.

7. Discussion and Conclusions

We have shown the available SpaceX Starlink satellite orbital plots as well as the sequence of events observed. The NORAD system was never able to identify 32 satellites. They were presumably lost between a few hours to days after launch. This implies quite heavy drag in the equatorial to midlatitude (up to 53°latitude) regions of the atmosphere at ~200 km altitude. At the present time there is not a known mechanism to cause such strongly enhanced drag at such low latitudes and altitudes. This will be a priority to investigate the physics and causes of this effect during magnetic storms.

Some of the satellites did survived the dual storm event. Since all the Starlink satellites were launched at the same time and at the same altitude, and they had such widely varying fates (some being immediately lost, some surviving) it is clear that each one had a different response to the magnetic storm density effects and/or had strong collisions with other satellites.

It took several more days for NORAD to make available the tracking of another train of 11 other satellites. The latter satellites were in more favorable positions (altitudes), allowing their recovery and rise to more stable orbits.

One can note from the above discussion that different satellites had extremely different orbital decay rates, indicating that one scenario can not fit all 43 satellite cases. In particular, we are most concerned about the possible losses of 32 of the satellites within the first 48 hours of launch such that they could never be tracked by NORAD.

8. **Comments on Previous Explanations of Starlink Satellite Losses**

Before it was known that the Starlink satellites never reach their intended ~500 km altitude, Tsurutani et al. (2022) proposed that prompt penetrating electric fields (PPEFs; Tsurutani et al., 2004; Tsurutani et al., 2007; Lakhina and Tsurutani, 2017) could be responsible for those losses. Their Figure 2 (reshown here as Figure 2) demonstrated that dayside near-equatorial density increases occurred at 500 km altitude during the two magnetic storms. However, the present orbital analyses indicate that none of the satellites lost on the first two days reached altitudes higher than 200 km for the entire orbit (they were still in elliptic trajectories). Thus, this loss mechanism must be discarded for the Starlink cases. However, on a positive note, it was shown for the first time using the Swarm satellite deceleration data that storm time PPEFs may be a main loss mechanism for satellites orbiting at ~400 to 500 km altitudes.

On the other hand, the Dang et al. (2002) scenario does not explain completely such losses in so low latitudes. They used a global upper atmospheric model (TIEGCM) to estimate the Joule heating by Ohmic dissipation at ionospheric altitudes. However, the Joule heating proposed by the authors was more remarkable in high latitudes, while the increases observed in latitudes below 53° were too small to create such an effect. Dang et al. predicted losses in 5 to 7 days assuming a constant 210 km satellite altitude. This cannot explain the possible immediate losses of the 32 satellites.

Fang et al. (2022) have used numerical simulations to show 50-125% neutral density enhancements between 200 and 400 km. Their argument based on effects of Joule heating produced in high latitudes propagating to lower latitudes by large-scale gravity waves, with phase speeds of 500 to 800 m/s (Fuller-Rowell et al., 2008). This propagation may take from 3 to 4 hours and are in addition to the effects of increased UV and EUV fluxes due to the flares. Previous events had taken up to 30 hours to the atmosphere returns to the undisturbed condition. We note, however, that Figure 2 in the present work showed that there were very low Joule heating effects in the auroral zone during both of these magnetic storms, thus negating the high latitude Joule heating effects assumed in the model.

9. **Final Comments**

After one year, with the small amount of tracking data made public, and the reduce number of satellites tracked in this mission, it is still not possible to clearly identify the true cause of the drag increases that lead to such losses. We thus suggest that researchers continue to try to determine the Starlink satellite orbits and consider other/new loss mechanisms.

## Acknowledgments

The work of R.H. is funded by the Chinese Academy of Sciences "Hundred Talents Program". E.E. would like to thank Brazilian agency for research grant: CNPq (contract no.PQ-301883/2019-0).